# The role of metallic impurities in oxide semiconductors: first-principles calculations and PAC experiments

**L. A. Errico**[*], **G. Fabricius,** and **M. Rentería**

Departamento de Física, Facultad de Ciencias Exactas, Universidad Nacional de La Plata, CC 67, 1900 La Plata, Argentina

We report an *ab-initio* comparative study of the electric-field-gradient tensor (EFG) and structural relaxations introduced by acceptor (Cd) and donor (Ta) impurities when they replace cations in a series of binary oxides: $TiO_2$, $SnO_2$, and $In_2O_3$. Calculations were performed with the Full-Potential Linearized-Augmented Plane Waves method that allows us to treat the electronic structure and the atomic relaxations in a fully self-consistent way. We considered different charge states for each impurity and studied the dependence on these charge states of the electronic properties and the structural relaxations. Our results are compared with available data coming from PAC experiments and previous calculations, allowing us to obtain a new insight on the role that metal impurities play in oxide semiconductors. It is clear from our results that simple models can not describe the measured EFGs at impurities in oxides even approximately.

## 1  Introduction

The experimental study of nuclear-quadrupole interactions is often used as a powerful tool to obtain information about local symmetry, coordination, and valence of defect or structural centers in solids. In the case of pure electric-quadrupole interactions, the measured quantities are the quadrupole coupling constant $v_Q = eQV_{33}/h$ and the asymmetry parameter $\eta = (V_{11} - V_{22})/V_{33}$, where $V_{ii}$ are the components of the diagonalized electric-field gradient (EFG) tensor ($|V_{33}|>|V_{22}|>|V_{11}|$). The EFG is measured via its interaction with the nuclear-quadrupole moment $Q$ of a suitable probe-atom (generally an impurity in the system under study) by different techniques, such as Perturbed-Angular Correlations (PAC). The presence of the impurity probe-atom modifies the local electrostatic potential and creates its own characteristic EFG.

As there is not a simple and accurate model to obtain the charge distribution from known experimental values of the $V_{ii}$ components, conclusions drawn from the experiments are usually restricted to symmetry considerations and empirical trends. Approaches for a quantitative calculation of $V_{ii}$ are often based on the point-charge model (PCM). Such an interpretation of the EFG measurements depends on the applicability of anti-shielding factors, it could only provide crude information on a possible charge transfer (ionicity) and ignores covalency. Therefore, it is desirable to calculate the EFG in an *ab initio* approach. Since the EFG is very sensitive to the anisotropic charge distribution close to the probe-nucleus, for its accurate calculation the entire electronic configuration of the host, perturbed by the presence of the impurity, has to be determined. This can be done in the frame of the density-functional theory. In 1999, we began a systematic study using the Full-Potential Linearized-Augmented Plane Waves (FLAPW) method

---

[*] Corresponding author: e-mail: errico@fisica.unlp.edu.ar, Phone: +54 - 221 - 4839061, Fax: +54 - 221 - 4252006

of electronic and structural properties at impurities in oxides, starting with Cd in $TiO_2$ and $SnO_2$ [1–3] ($^{111}$Cd is widely used as PAC probe). In this work we present results at the Ta impurity in $TiO_2$, $SnO_2$, and $In_2O_3$ ($^{181}$Ta is the second widely used PAC probe) and we make a comparative study with the results previously obtained for Cd. This approach gives very good results in doped systems and is not only able to calculate structural and electronic properties of the systems accurately, but also to provide a deeper understanding of them.

## 2   The systems under study

Rutile $TiO_2$ and $SnO_2$ are tetragonal ($a = b = 4.5845_1$ Å, $c = 2.9533_1$ Å for $TiO_2$ [4] and $a = b = 4.7374_1$ Å, $c = 3.1864_1$ Å for $SnO_2$ [5]), the unit cells (Fig. 1) contains 2 cations at (0, 0, 0) and (1/2, 1/2, 1/2) and 4 O at $\pm(u, u, 0; \frac{1}{2} + u, \frac{1}{2}-u, \frac{1}{2})$ with $u = 0.30493_7$ and $0.3056_1$ for $TiO_2$ and $SnO_2$, respectively [4, 5]. In order to simulate the isolated impurities we used super-cells (SC) of 12 unit-cells with one cation replaced by the Cd/Ta atom. The resulting 72-atoms SC has dimensions $a' = b' = 2a$, $c' = 3c$.

## 3   Results and discussion

We have calculated from first-principles the EFG tensor at Cd/Ta impurities replacing cations in $TiO_2$ and $SnO_2$ taking into account the structural relaxations introduced by the impurities in the host lattices. To deal with this problem we performed FLAPW calculations with the WIEN97 code [6] within the LDA [7] approximation. We took for the parameter $RK_{MAX}$, which controls the size of the basis-set, the value of 7 and we introduced local orbitals to include Ti-3$s$ and 3$p$, Sn-4$d$ and 5$p$, O-2$s$, and Ta-6$s$, 5$p$, and 4$f$ orbitals. Details of the method of calculation and the way to deal with the impurity are extensively explained in Ref. [2].

### 3.1   Cd and Ta in $TiO_2$

There is an important point to be considered in the calculation of the electronic structure of the systems: the charge state of the impurity-host system. In this paragraph we discus the case of the double acceptor $Cd^{+2}$ in $Ti^{+4}O_2^{-2}$ (described in detail in Ref. [2]) in order to compare afterwards with the Ta results. $TiO_2$ is a semiconductor with the oxygen $p$-band filled. When a Cd atom replaces a Ti in the SC, the resulting system is metallic due to the lack of two electrons necessary to fill the oxygen $p$-band. Comparison of Fig. 2a and 2b shows that the presence of Cd in the SC produces the appearance of Cd-$d$ levels and impurity states at the top and the bottom of the valence band. The wave function of the impurity state at the Fermi level ($E_f$) has character Cd-$d_{yz}$, $O_1$-$p_y$, $O_1$-$p_z$. Then, to provide two electrons to the system implies a drastic change in the symmetry of the electronic charge distribution in the neighborhood of the impurity. That is why we have found different relaxations and EFGs for different charge states of the impurity. We show results assuming that: a) extra electrons are not available (neutral impurity, $Cd^0$); b) the system provides the lacking two electrons (charged impurity, $Cd^{-2}$). In situation (a) we used the described SC. To describe situation (b) we added two electrons to the SC that we compensated with an homogeneous positive background to obtain a neutral SC to compute energy and forces. In Table 1 we show the results for the relaxation of the six nearest oxygen-neighbors (O$_{NN}$) to the Cd atom. We see that for both charge states the relaxations are quite anisotropic, with the Cd-O$_1$ distance larger than the Cd-O$_2$ distance, opposite to the un-relaxed structure. This result contradicts the assumptions of previous studies [8, 9]. The change in the local symmetry around Cd produces a change in the orientation of $V_{33}$ from the [001] to the [110] direction upon substitution of a Ti atom by a Cd impurity. The results for the EFG tensor for Cd$^{-2}$ agree very well with the experimental results (see Table 1) and are very different from the ones obtained for Cd$^0$. This difference in the EFG is originated in the filling of the impurity state at $E_F$, which has Cd-$d_{yz}$ character. The high $\eta$ value obtained for Cd$^0$ is too far from experimental data showing that Cd in TiO$_2$ is in a charged state. Finally, the results obtained changes very slightly when ions beyond the O$_{NN}$ are relaxed.

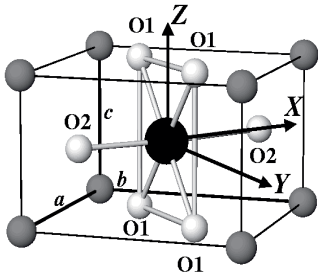

**Fig. 1**   Unit cell of rutile $TiO_2$ and $SnO_2$ (cations gray balls, O white balls). The results discussed in this work are referred to the indicated axes system, assuming that the impurity (black atom) replaces the central cation.

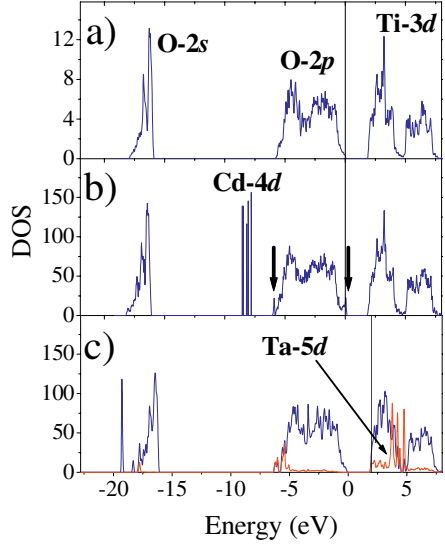

**Fig. 2** DOS for a) pure TiO$_2$; b) Cd$^0$ in TiO$_2$ (72 atoms SC). The arrows indicate impurity states in the valence band; c) Ta$^0$ in TiO$_2$ (72 atoms SC). The amplitude of the Ta-5$d$ DOS was multiply by 20 for a better visualization. The vertical lines indicate the Fermi level.

**Table 1** Final distances of the O$_{NN}$ and EFGs for Cd and Ta in TiO$_2$ and SnO$_2$ in units of $10^{21}$ V/m$^2$. $d(m – O_1)$ and $d(m – O_2)$ are the distances in Å (after relaxation) from the impurity $m$ to O$_1$ and O$_2$ atoms, respectively ($m$: Cd, Ta). Sign of the experimental $V_{33}$ values are unknown. In first row, distances and EFG tensor refer to Ti site in TiO$_2$.

| System | | $d(m – O_1)$ | $d(m – O_2)$ | $V_{XX}$ | $V_{YY}$ | $V_{ZZ}$ | $V_{33}$ | $V_{33}$ direction | $\eta$ |
|---|---|---|---|---|---|---|---|---|---|
| TiO$_2$ exp. [13] | | 1.94 [4] | 1.98 [4] | | | | 2.2(1) | Z | 0.19(1) |
| TiO$_2$:Cd | Cd$^0$ [1, 2] | 2.15 | 2.11 | –7.16 | +6.82 | +0.34 | –7.16 | X | 0.91 |
| | Cd$^{2-}$ [1, 2] | 2.18 | 2.11 | –2.87 | +4.55 | –1.68 | +4.55 | Y | 0.26 |
| | PCM | | | –2.21 | +1.58 | +0.63 | –2.21 | X | 0.43 |
| | Exp. [1] | | | | | | 5.34(1) | X or Y | 0.18(1) |
| TiO$_2$:Ta | Ta$^0$ | 1.95 | 1.96 | +2.0 | +10.0 | –12.0 | –12.0 | Z | 0.66 |
| | Ta$^+$ | 1.95 | 1.97 | +2.0 | +11.0 | –13.0 | –13.0 | Z | 0.69 |
| | PCM | | | –4.56 | +3.26 | +1.30 | –4.56 | X | 0.43 |
| | Exp. [11] | | | | | | 13.3(1) | Z | 0.56(1) |
| SnO$_2$, FLAPW [12] | | 2.06 [5] | 2.05 [5] | +5.56 | –1.15 | –4.41 | +5.56 | X | 0.59 |
| SnO$_2$:Cd | Cd$^{-2}$ [3] | 2.20 | 2.16 | –2.94 | +5.35 | –2.41 | +5.35 | Y | 0.10 |
| | PCM | | | –3.65 | +5.22 | –1.57 | +5.22 | Y | 0.40 |
| | Exp. [13] | | | | | | 5.83(4) | – | 0.18(2) |
| SnO$_2$:Ta | Ta$^+$ | 1.98 | 1.99 | +0.1 | +17.6 | –17.7 | –17.7 | Z | 0.96 |
| | PCM | | | –7.53 | +10.76 | –3.23 | +10.76 | Y | 0.40 |
| | Exp. [14] | | | | | | 17.04(1) | – | 0.70(2) |

Ta in TiO$_2$ is a simple donor impurity. We study two charge states: neutral impurity state, Ta$^0$, and charged impurity state, Ta$^+$. Contrary to Cd, the results for the structural relaxations and the EFG are independent of the charge-state of the impurity (see Table 1). This fact can be understood looking at Fig. 2: the Ta-$5d$ levels are located at the same energy of the Ti-$3d$ levels, and impurity levels with a particular symmetry do not appear at the bottom edge of the conduction band. Another difference between Cd and Ta is the magnitude of the relaxations: the relaxations introduced in TiO$_2$ by Ta are much smaller than those of Cd (see Table 1). For this reason there is no change of the orientation of $V_{33}$ with respect to pure TiO$_2$. Besides the small magnitude of the relaxation, it has to be taken into account for the description of the EFG, since at the un-relaxed structure we obtain $\eta = 0.0$, in bad agreement with experiments. Finally, the agreement between FLAPW and PAC results is very good.

### 3.2 Cd and Ta in SnO$_2$

We will discus the results obtained for the impurities in their charged state (Cd$^{-2}$ and Ta$^+$). In the case of Cd (see Table 1) our results show that, similar to Cd in TiO$_2$, the relaxation is not isotropic. It is important to note that the final distances Cd-O$_{NN}$ in SnO$_2$ and TiO$_2$ are similar. It seems that the local structure try to reconstruct the environment of Cd in its own oxide, CdO (distances Cd-O$_{NN}$ are rather large in this oxide, c.a. 2.35 Å). With regard to the EFG, the results obtained for $V_{33}$ and $\eta$ after the O$_{NN}$ relaxation reproduce successfully the experimental values.

In the case of the Ta impurity, the relaxation (contraction) of the O$_{NN}$ is small. This result is similar to the one found for Ta in TiO$_2$ and can be understand from the fact that the bond-lengths of Ta in rutile TaO$_2$ is about 2.02 Å. Again, the local structures try to reconstruct the environment of Ta atoms in its own oxide. This assert is in agreement with FLAPW results at Ta located at cationic site $D$ of In$_2$O$_3$ that predict a contraction of the Ta-O$_{NN}$ distances from 2.20 Å (pure In$_2$O$_3$) to 2.04 Å. Concerning the EFG, the FLAPW results are in very good agreement with the experimental ones (see Table 1).

The simplest and widely used approximation for the calculation of the EFG at a probe-atom is the PCM. In both oxides and for both impurities, this model fails in the prediction of the EFGs (see Table 1). Even if the relaxed coordinates from our FLAPW calculations are introduced in the PCM, this model still fails in the description of the EFGs [12]. If we compare this traditional view with our FLAPW results and the experimental values, we arrive to a completely different understanding of the EFG: based on the PCM, one expects that the lattice gives an EFG that is amplified by the Sternheimer anti-shielding factor $\gamma_\infty$. In Table 1 we can see that PCM predictions are very different for TiO$_2$ and SnO$_2$, but after the relaxations, FLAPW and the experiments show similar results for both compounds. This fact shows that the impurity, and the electronic and structural process that it induces, mainly determine the EFG.

To conclude, we want to comment preliminary FLAPW results obtained for Cd at both cationic sites in In$_2$O$_3$. Because the bonding in this oxide is mainly ionic, it has been usually supposed that PCM is a good model to evaluate the EFG at both sites [15, 16]. We found that, contrary to previous suppositions [15], Cd induces relaxations of about 5% of the Cd-O$_{NN}$ bond-lengths with respect to the un-relaxed structure. The FLAPW prediction for the EFG at Cd impurities at site $D$ ($V_{33} = +7.6 \times 10^{21}$ V/m$^2$; $\eta = 0.0$) and site $C$ ($V_{33} = +5.6 \times 10^{21}$ V/m$^2$; $\eta = 0.68$) are in excellent agreement with the PAC results ($V_{33} = 7.7_1 \times 10^{21}$ V/m$^2$; $\eta = 0.05_5$ (site $D$), $V_{33} = 5.9_1 \times 10^{21}$ V/m$^2$; $\eta = 0.69_2$ (site $C$) [17]). While PCM and FLAPW are in agreement for site $D$, this is not the case for site $C$ where PCM predicts an opposite sign ($V_{33} = -4.9 \times 10^{21}$ V/m$^2$).

## 4  Conclusions

Ab initio calculations of the EFG by the FLAPW method successfully predicted the EFG at Cd/Ta impurity sites in TiO$_2$, SnO$_2$, and In$_2$O$_3$ yielding quantitative information on the charge state of the impurities and the lattice relaxation that it induces. From our results it is clear that the problem of the EFG at impu-rities in oxide semiconductors is too complicated to be described (even approximately) by simple models like PCM, antishielding factors, and arbitrary suppositions like isotropic relaxations. We can conclude that a proper theoretical description of electronic properties of metal impurities in these systems should consider self-consistently the charge-state of the impurity and the impurity-induced distortions in the hosts, especially in the first shell of neighbors of the impurity.

**Acknowledgements** This work was partially supported by CONICET, Fundación Antorchas, and ANPCyT (PICT98 03-03727), Argentina, and TWAS, Italy. L. A. E. is fellow of CONICET. G.F. and M.R. are members of CONICET.